\documentstyle[11pt,newpasp,twoside]{article}
\begin{document}
\hyphenation{New-to-ni-an}
\hyphenation{post-New-to-ni-an}

\title{Calculating Gravitational Radiation from Collisions}
 \author{Joan M. Centrella}
\affil{Department of Physics, Drexel University, Philadelphia, PA  19104}

\begin{abstract}
The development of both ground- and space-based gravitational
wave detectors provides new opportunities to observe the radiation
from binaries containing neutron stars and black holes.  Numerical
simulations in 3-D are essential for calculating the coalescence
waveforms, and comprise some of the most challenging problems in
astrophysics today.  This article briefly reviews the current status
of efforts to calculate black hole and neutron star coalescences, and
highlights challenges for the future.
\end{abstract}

\section{Introduction}

Binaries containing black holes (BHs) or 
neutron stars (NSs) are among the most
important and interesting sources for the gravitational wave detectors
expected to start operation in the early part of the $21^{\rm st}$
century.  These binaries spiral together due to the emission of
gravitational radiation, leading to the final collision and
coalescence of their components.  The NS/NS, NS/BH, and stellar BH/BH binaries
are target sources for ground-based interferometers such as LIGO,
VIRGO, GEO600, and TAMA, while the 
space-based LISA is expected to be most sensitive to massive 
BH binaries (see the articles by Weiss and Kalogera in
these Proceedings). Astrophysically, the data from these
detectors can provide important insights
into NS properties; the equation of state of matter at nuclear
densities; models of gamma-ray bursts; active galactic nuclei and
quasars; and the strong-field regime of
gravity, including unambiguous detection of BH formation 
(Schutz 1997; Thorne 1998).

Calculations of the gravitational waves from such
binaries focus on three
physical regimes.  During the {\em inspiral phase}, the components
are widely separated and can be treated as point particles, allowing
analytic calculations using post-Newtonian (PN) expansions.  Templates
constructed from the resulting waveforms are expected to form a key
component of data analysis and source identification
(Flanagan \& Hughes 1998).
 The {\em coalescence phase} begins when
the compact objects are close enough to suffer tidal deformation,
and continues as a combination of relativistic and hydrodynamic
effects drives the stars together on more rapid timescales.
Calculation of the waveforms produced during this stage requires 3-D
numerical simulations that solve the full Einstein equations and, for
a NS, general relativistic hydrodynamics; in some
cases, the formation of a BH must also be modeled.  Finally, the
{\em ringdown phase} encompasses the late-time behavior of the merger
product or remnant, radiating in its normal modes (Thorne 1998).

Numerical simulation of the coalescence phase using full
 general relativity
is one of the most challenging problems in astrophysics
today.  This article surveys the current status of efforts to
calculate coalescence waveforms for NS/NS, NS/BH, and BH/BH binaries.

\section{Requirements and Challenges for Successful Simulations}

 The difficulties in the 3-D numerical simulation of binary
coalescence begin in specifying the initial state of the
system, which must correspond to a physically realistic
binary evolving through the PN inspiral phase
(Brady, Creighton, and Thorne 1998).  Specifically, there
should be no incoming gravitational radiation in the system, and both
the outgoing gravitational waves and the orbital parameters should
correspond to those produced during the PN inspiral stage.  These
conditions must all be met within the constraints of the initial value
Einstein equations, which are typically a set of coupled nonlinear
equations.  Although
relativistic simulations to date have used highly
simplified initial conditions that do not fulfill these requirments,
progress is being made towards developing more plausible initial
conditions (Alvi 2000; Buonanno \& Damour 2000; Damour, Jaranowski,
\& Sch\"{a}fer 2000; Marronetti \& Matzner 2000).

Once the initial data has been specified, the remaining Einstein
equations are solved to evolve the system in time.  
This evolution
must be stable over timescales corresponding to at least a few binary
orbits.  Various formulations of the Einstein equations, notably the
hyperbolic (Reula 1998; Alcubierre, et al. 1999)
 and conformal ADM (Shibata \& Nakamura 1995; Baumgarte \& Shapiro 1998)
 approaches, have been introduced in
recent years to improve the stability of the 
evolution\footnote{These formalisms assume that the initial data is
given on a 3-D spacelike Cauchy surface. Methods employing
characteristic or null slicings of spacetime are also under
development (Bishop, et al. 1999).}.  Additional
equations enter to specify the time-slicing of the spacetime and the
evolution of the spatial coordinates.  This latter gauge condition is
likely to be quite important in keeping the binary in a co-rotating
frame of reference; see below.  In the case of a NS or WD, the
equations of general relativistic hydrodynamics must also be solved,
with high-resolution shock-capturing techniques playing an increasingly
important role (Font, et al. 2000).
  If the binary contains a BH, additional
challenges arise in moving the BH across the grid.  While
current efforts
focus on excising the interior of the BH, which avoids the central
singularity, the long-term stability of the inner
 boundary at the apparent horizon
remains a challenge (Alcubierre \& Bruegmann 2000).
  And, the code must be capable of handling the
formation of a BH from the merger, an event that is generally detected
numerically by
finding an apparent horizon around the remnant (Alcubierre, et
al. 2000; Huq, Choptuik, \&
Matzner 2000).  The code should
continue to evolve stably as the system enters the ringdown phase.

The gravitational radiation emitted during the coalescence produces
its own set of challenges.  Typically, the radiation will have
wavelengths
$\lambda_{\rm GW} \sim 10 - 100 L$, where $L$
 is the scale of the source.  The computational domain must be
large enough to allow the signals produced near the source to
propagate outward and develop into gravitational waves in the
radiation zone, and at the same time provide adequate resolution for
the source dynamics.  Such widely varying length and time scales point
to the use of adaptive mesh refinement (AMR) as a critical component
of success (Papadopoulos, Seidel, \& Wild 1998; New, et al. 2000).

Gravitational wave detectors are located asymptotically far
from the astrophysical sources.  For a meaningful comparison with the
observations, the radiation generated in the simulations must also be
measured ``at infinity.''  One promising means of achieving this is to
extract the waves on a spacelike slice at some distance 
$\gg L$, and then propagate them along the characteristics to
infinity using a different code (Bishop, et al. 1998).
  Of course, there will be some
scattering of these waves back towards the source and thereby back
into the original code.  Currently, efforts are aimed at
achieving this matching in a stable manner,
and thereby producing good, physical outer boundary conditions for the
source evolution codes.

Finally, numerical relativists generally use high performance
computers and sophisticated visualization techniques to perform their
simulations and extract the physics from the results. Further progress
depends on continued access to advanced computing resources
(Suen 1999).

\section{Simulations of NS/NS Coalescence}

Binaries containing 2 NSs have long been considered among the  most
promising sources for the first generation of ground-based
interferometers. Such instruments could observe the inspiral phase of
these systems during the last few minutes of their evolution,
as the gravitational waves sweep upward in frequency from 
$\sim 10$ Hz to $\sim 1000$ Hz, yielding several thousand cycles of
the signal across the broadband range of the detectors (Cutler, et al.
1993).  The
coalescence
phase is expected to occur at rather high frequencies, typically
$\ga 1$kHz. Observations of coalescence are of
considerable interest,
as the waveforms are expected to contain information
about NS structure and the nuclear equation of state at high densities
(Zhuge, Centrella, \& McMillan 1994, 1996), as well as general
relativistic effects. Coalescing NS binaries are also astrophysically
important as models for $\gamma-$ray bursts (Ruffert, Janka, \&
Sch\"{a}fer 1996) and in the production of r-process nuclei
(Rosswog et al. 1998).  

Today, there are a variety of NS/NS coalescence simulations, ranging 
from those using the Newtonian limit to models with full general
relativity. Simulations have been carried out using both synchronized
(co-rotational) and non-synchronized (irrotational) binaries.  The latter are
considered more realistic, since NS viscosities are expected to be too
small to cause synchronization before the stars coalesce (Kochanek
1992; Bildsten \& Cutler 1992). In this article we will give a brief
survey of coalescing NS simulations,
highlighting important lessons that have been learned.  For a more
complete list, see the recent review by Rasio \& Shapiro (1999).

\vspace{0.07in}
\noindent
{\em Newtonian Models}
\vspace{0.03in}

In models with purely Newtonian gravitational fields, the
gravitational waves are calculated using the quadrupole approximation
(Misner, Thorne, \& Wheeler 1973) with no gravitational
back-reaction.  Researchers in this area have used either grid-based
Eulerian hydrodynamics or Lagrangian smoothed particle
hydrodynamics (SPH); see Faber, Rasio, \& Manor
(2000) for references.
Such models are strictly valid only in the limit
$GM/Rc^2 \ll 1$, for an object of mass $M$ and radius $R$.  Since
$GM/Rc^2 \sim 0.2$ for a typical NS, Newtonian simulations
can only provide a rough guide to the physics expected in
fully relativistic models of NS/NS
coalescence.  Nevertheless, Newtonian simulations remain
an important tool for exploring gravitational wave phenomena and
provide a simplified arena in which numerical techniques can be
tested.  

Coalescence in the Newtonian limit is driven by global hydrodynamic
instabilities (Lai, Rasio, \& Shapiro 1994; Rasio \& Shapio 1992) that
cause the stars to plunge towards each other once they are close enough
for tidal effects between them to become sufficiently strong. SPH
simulations by Zhuge, Centrella, \& McMillan (1994, 1996) in which the
NSs are modeled as polytropes show that the gravitational wave energy
spectrum $dE/df$ has a steep drop at a cutoff frequency, typically
$f_{\rm cut} \sim 2$kHz. The exact location of $f_{\rm cut}$ depends
on the NS radius $R$, the stiffness of the equation of state, the NS
spin, and the mass ratio of the components.  Ringing oscillations of
the merged remnant can produce additional structure at even higher
frequencies. 
	New \& Tohline (1997) carried out Eulerian simulations of
synchronized binaries in both rotating and inertial (non-rotating)
reference frames.  They showed that the numerical viscosity caused by
fluid flowing through the grid can produce spurious instability
leading to coalescence.  Thus a binary that is stable when evolved in
the rotating frame, becomes unstable to merger when computed in the
inertial frame.  A recent study by Swesty, Wang, \& Calder (1999)
further underscores the importance of using a rotating frame, and also
explores the effects of solving for the gravitational field with the
density at the previous time step (time-lagged), the new time step
(time-advanced), or an average of the old and new values
(time-centered).  With gridless SPH models, Rosswog, et al.(2000)
have shown
the effects of various 
prescriptions for artificial viscosity in causing coalescence
of NS/NS binaries.

\vspace{0.07in}
\noindent
{\em Post-Newtonian Models}
\vspace{0.03in}

The next level of sophistication incorporates the lowest order
post-Newtonian corrections to Newtonian gravity (1PN) and
gravitational radiation reaction (2.5PN).  Early calculations were
carried out by Oohara \& Nakamura (1992) using Eulerian techniques.
The most recent PN simulations have been done by Faber, et al. (2000;
see also Faber \& Rasio 2000) using SPH, who find that the addition of
1PN corrections can lower the maximum gravitational wave luminosity
and produce differences in the shape of the signal compared to
Newtonian models. 

The use of PN corrections in NS/NS simulations poses certain
problems.  For consistency, all 1PN quantities must be small,
requiring $GM/Rc^2 \ll 1$.  To this end, Shibata, Oohara, \& Nakamura
(1997) and Ayal, et al. (1999) carried out 1PN mergers using $GM/Rc^2
\sim 0.03$.  However, a realistic NS typically has $GM/Rc^2 \sim
0.2$. Also, since the 2.5PN radiation reaction terms scale as
$(GM/Rc^2)^{2.5}$, imposing $GM/Rc^2 \ll 1$ results in very weak
radiation reaction effects.  Faber, et al. (2000) employ a hybrid
scheme in which realistic NS parameters are used for the 2.5 PN
radiation reaction terms, and the 1PN terms are scaled down
artificially to keep them small.

Additionally, the effects of PN corrections on the separation at which
the NSs leave their quasi-circular inspiral orbits and begin to plunge
toward the center are not yet clear.  Faber \& Rasio (2000) find that
the PN corrections cause the plunge to begin at larger separations
than in the Newtonian limit.  However, Shibata, Taniguchi, \& Nakamura
(1998) have analyzed equilibrium sequences of binary stars, and find
that the orbits generally become unstable at smaller separations as PN
corrections are increased.  This is an important issue to resolve,
since a plunge beginning at larger separation will shift the
coalescence waveform to lower frequencies.  Newtonian simulations show
the coalescence occurring at relatively high frequencies that are
outside the broad-band sensitivity of first-generation interferometers
such as LIGO-1 (Zhuge, et al. 1994, 1996).  New experimental
techniques may improve the sensitivity of advanced interferometers at
high frequencies (Meers 1988, Strain \& Meers 1991; see also
http://www.ligo.caltech.edu/$\sim$ligo2/).

\vspace{0.07in}
\noindent
{\em General Relativistic Models}
\vspace{0.03in}

The earliest simulations of binary NS coalescence in full general
relativity were carried out by Nakamura, Oohara, \& Kojima (1987)
using Eulerian techniques.  Although there are currently a number of
efforts worldwide aimed at developing relativistic coalescence models
(e.g. Font, et al. 2000; Landry \& Teukolsky 1999; Baumgarte, et al.
1999)
this subject is still in its infancy.  All fully relativistic codes to date
are Eulerian.

The most advanced relativistic coalescence simulations have been
carried out by Shibata \& Ury\={u} (2000).  They solve the full
Einstein equations using a conformal ADM formalism; these are coupled
to the hydrodynamic equations that govern the sources, and the  matter
is taken to be a perfect fluid (Shibata 1999). Their initial data
consists of 2 identical NSs, modeled as polytropes with $\Gamma=2$, in
a quasi-equilibrium orbit.  Both irrotational and corotational models
are considered.  The orbit is then destabilized by reducing the
angular momentum from its quasi-equilibrium value, and merger ensues.
They find that a BH is formed on a dynamical timescale if the initial
NSs are sufficiently compact.  For corotational binaries, the remaining
disk of matter around the BH can have a mass $\sim 5\% - 10\%$ of the
total system rest mass.  The merger of irrotational NSs to form a BH
produces a disk with a much smaller mass, $< 1\%$ of the total system
mass.  For less compact initial NSs, the result of the coalescence is a
differentially rotating massive NS. 

\section{Simulations of NS/BH Coalescence}

Coalescence of a binary containing a NS and a stellar BH is another
important source for ground-based gravitational wave detectors.
Recent work suggests that observations of tidal disruption of a NS by
a BH may provide measurements of the NS radius and equation of state
(Vallisneri 2000).  NS/BH systems are also astrophysically interesting
as engines for $\gamma-$ray bursts (Janka, et al. 1999) and in the
production of r-process nuclei (Lee 2000).

Simulations of NS/BH coalescence have been carried out in the
Newtonian limit by Lee \& Klu\'{z}niak (1995; 1999a,b; Klu\'{z}niak 1998)
using SPH and by Janka, et al. (1999) using Eulerian techniques.  In
these efforts, the BH is taken to be a Newtonian point mass surrounded
by an absorbing boundary at the Schwarzschild radius.  These Newtonian
codes are also extended to include gravitational wave emission and
backreaction using various techniques.  Lee \& Klu\'{z}niak modeled
the NS as a polytrope, whereas Janka, et al. used a more physical
description of NS matter.

The most recent NS/BH coalescence models have been calculated by Lee
(2000) using SPH with irrotational initial conditions.  He finds that
the fate of the system depends strongly on the stiffness of the
equation of state.  After an initial period of mass transfer, the NS
is completely disrupted by the BH during its second periastron passage
for polytropic index $\Gamma = 2.5$. However, for a stiffer model with
$\Gamma = 3$, the NS is not completely disrupted; instead, a remnant
is left in a higher orbit about the BH.  The resulting gravitational
wave signals clearly reflect the different outcomes. 

Of course,  
relativistic effects are expected to be very important in NS/BH
systems.  For this reason, simulations incorporating full general
relativity may introduce qualitatively new features in the waveforms
and are eagerly awaited.

\section{Simulations of BH/BH Coalescence}

The merger of BH/BH binaries is, in many ways, the quintessential
general relativistic source of gravitational radiation.  Binaries
containing BHs with masses $\sim 10M_{\sun}$ are an important target for
ground-based interferometers, while the coalescence of binaries
containing  massive BHs
$\sim 10^5 - 10^6 M_{\sun}$ is a prime candidate for
observation by LISA. Since the dynamics of BH/BH binaries scales with
the total system mass $M$, observations in each of these mass ranges
are pertinent to the same basic BH physics (Thorne 1998).  In
particular, the gravitational waves produced during these mergers are
expected to have large amplitudes and highly nonlinear 
characters, and to provide important tools for understanding dynamical
spacetime curvature.  Astrophysically, the detection of BH/BH
coalescences can yield important information on 
BH binary  formation in
dense stellar systems such as galactic nuclei and globular clusters
(Portegies Zwart \& McMillan 2000), as well as on active galactic
nuclei and binary quasars (Mortlock, Webster, \& Francis 1999).

Numerical simulations of BH/BH coalescence are, by nature, fully
general relativistic.  The early work of Hahn \& Lindquist
(1964) was followed by a major effort by Smarr and Eppley (Smarr 1978;
Eppley 1977) that produced the first simulations of gravitational
radiation from axisymmetric, head-on collisions of equal mass
BHs in 2-D.  Smarr and Eppley pioneered the use of
singularity-avoiding spacetime slicing in numerical relativity, which
was used throughout the 1980s and 1990s.  The most
recent work using this approach is the simulation of non-head-on
collisions of spinning BHs in 3-D by Bruegmann (1999).

However, singularity-avoiding slicings typically give rise to
problems,
such as grid stretching, 
that cause the simulations to fail after a relatively short time.  This
technique is thus unsuitable for modeling BH/BH binaries for long
durations, such as the time from the
late inspiral phase (when the point-mass approximation breaks down)
through coalescence and ringdown.  A more promising approach is to use
techniques in which the interior of the BH is excised from the
computational domain (Seidel \& Suen 1992).  Current efforts are
focused on developing appropriate ``inner boundary conditions'' at the
excision surface that will allow long-term, stable evolutions
(Alcubierre \& Bruegmann 2000; Lehner, et al. 2000).  

The first
simulations of non-head-on BH collisions using singularity excision
were recently carried out by Brandt, et al. (2000), with both spinning
and non-spinning BHs.  They used apparent horizon conditions to mark
the excised regions, and were able to follow the BHs through merger,
signalled by the formation of a single apparent horizon around the
system.  Preliminary estimates of the area of this horizon indicate
that $\sim 2.6\%$ of the total mass is radiated away as gravitational
waves.
After a moderate time, however, instabilities caused these
models to fail.  Efforts are underway to determine the source of these
instabilities, and to develop techniques to eliminate them.

\section{Outlook}

Gravitational wave astronomy offers many exciting opportunities to
explore the physical universe.  Significant challenges remain,
however, in the experimental, theoretical, and computational arenas.
In particular, fully general relativistic simulations are currently in
their infancy; while important pieces of the puzzle are starting to
take shape, considerable effort will be required to produce robust,
reliable models.  Ultimately, success in gravitational wave astronomy
can be expected to involve partnerships and collaborations among
experimentalists, relativity theorists, computational modelers,
astronomers, and astrophysicists.  The resulting synergy and
creativity promise a bright and exciting future.

\bigskip
\acknowledgments
It is a pleasure to thank Kimberly New and Richard Matzner
for comments on the manuscript.
This work was supported by NSF grant PHY-9722109.

\end{document}